\documentclass[useAMS,usenatbib]{mn2e}
\usepackage{graphics,graphicx}
\newcommand{\SM}{M$_{\odot}~$}
\newcommand{\gadget}{\emph{Gadget-2}}
\newcommand{\msmbh}{M$_\bullet$}

\newcommand{\beq}{\begin{equation}}
\newcommand{\eeq}{\end{equation}}

\title[The Doubling of Stellar Black Hole Nuclei]{The Doubling of Stellar Black Hole Nuclei}
\author[Mher V. Kazandjian and Jihad R. Touma]{Mher V. Kazandjian$^{1}$\thanks{E-mail:
mher@strw.leidenuniv.nl; jt00@aub.edu.lb} and J. R. Touma$^{2}$\footnotemark[1]\\
$^{1}$Leiden Observatory, Leiden University, P.O. Box 9513, 2300 RA
 Leiden, The Netherlands\\
$^{2}$Department of Physics, American University of Beirut, Beirut,
 Lebanon}
\begin{document}

\date{Accepted yyyy mmmmmmmm dd. Received yyyy mmmmmmmm dd; in original form yyyy mmmmmmm xx}

\pagerange{\pageref{firstpage}--\pageref{lastpage}} \pubyear{2012}

\maketitle

\label{firstpage}

\begin{abstract} It is strongly believed that Andromeda's double
nucleus signals a disk of stars revolving around its central
super-massive black hole on eccentric Keplerian orbits with nearly
aligned apsides. A self-consistent stellar dynamical origin for such
apparently long-lived alignment has so far been lacking, with
indications that cluster self-gravity is capable of sustaining such
lopsided configurations if and when stimulated by external
perturbations. Here, we present results of N-body simulations which
show unstable counter-rotating stellar clusters around super-massive
black holes saturating into uniformly precessing lopsided nuclei. The
double nucleus in our featured experiment decomposes naturally into a
thick eccentric disk of apo-apse aligned stars which is embedded in a
lighter triaxial cluster. The eccentric disk reproduces key features
of Keplerian disk models of Andromeda's double nucleus; the triaxial
cluster has a distinctive kinematic signature which is evident in HST
observations of Andromeda's double nucleus, and has been difficult to
reproduce with Keplerian disks alone. Our simulations demonstrate how
the combination of eccentric disk and triaxial cluster arises
naturally when a star cluster accreted over a pre-existing and
counter-rotating disk of stars, drives disk and cluster into a
mutually destabilizing dance.  Such accretion events are inherent to
standard galaxy formation scenarios. They are here shown to double
stellar black hole nuclei as they feed them. \end{abstract}


\section{Introduction} Over the years, the  nucleus of the Andromeda
galaxy (M31) went from being asymmetric, to doubling, then
tripling. The asymmetry was first noted in the balloon-born
Stratoscope observatory \citep{Light1974}. It was photometrically
resolved into a double nucleus by Hubble Space Telescope (HST)
observations \citep{Lauer1993}; asymmetry hence turned into
lopsidedness, with a luminous feature (referred to as P1) shining a
few parsecs away from a dimmer one (referred to as P2), which is
closer to the center of the host bulge. The resolved double nucleus
had already been suspected to host a Super-Massive Black Hole (SMBH),
located somewhere close to P2, with a neighborhood that was known to
shine brighter than the rest of the nucleus in the ultraviolet
\citep{Dressler1988, Kormendy1988}. Detailed HST spectroscopy added a
third component (known as P3), a disk of young massive stars, whose
size and rotation speed rules out viable alternatives to a central
SMBH of $\sim 10^8$ solar masses {\it (\SM) } \citep{Bender2005}.

In the currently favored model of M31's lopsided double nucleus
\citep{Tremaine1995, Peiris2003}, the brighter peak P1
\citep{Lauer1993} is thought to coincide with the common apo-centric
region of an eccentric disc of stars revolving on nearly apse-aligned
Keplerian ellipses, around M31's central super-massive black hole
(hereafter SMBH). Similar eccentric discs are thought to underlie
lopsided nuclei detected around SMBHs in the centers of nearby
galaxies \citep{Lauer1996, Lauer2005, Gultekin2011}. That such
kinematic configurations can be sustained in self-gravitating disks
around SMBHs was clarified in a series of dynamical studies. Indeed,
investigation of modes of hot nearly-Keplerian disks indicated that
slow (m=1, lopsided) modes can be stably excited
\citep{Tremaine2001}. This conclusion was corroborated by results of
planar N-body simulations of disks around SMBHs which showed that,
when started in asymmetric conditions, disks can relax into long-lived
uniformly precessing lopsided configurations \citep{Bacon2001,
Jacobs2001}. Subsequently, razor-thin, self-gravitating, lopsided
equilibria were constructed, yielding encouraging agreement with
photometric and kinematic observations of M31's nucleus
\citep{Salow2001, Sambhus2002}. These and related studies left open
questions about how such modes are excited, how they ultimately
saturate into global lopsided configurations, and whether constructed
equilibria were stable or not.  The early suggestion
\citep{Tremaine1995} that an initially circular disk could become
eccentric under the influence of dynamical friction with the host
bulge has not been thoroughly explored. A recently proposed scenario
\citep{Hopkins2010a, Hopkins2010b} has eccentric disks forming in the
notoriously complex -and poorly understood \citep{Silk2011}-
environment which couples star formation and its feedback to gas
accretion in the early stages of SMBH growth.\\ \indent In this work,
we opt for a minimalist and dynamically self-consistent route to
three-dimensional, lopsided nuclei, out of counter-rotating (CR
hereafter) collisionless stellar distributions dominated by
SMBHs. Such CR distributions are prone to violent m=1 instabilities in
the presence of moderate counter-rotation \citep{Jog2009, Touma2002},
hence the suggestion that the lopsided structure in Andromeda's
nucleus may have been triggered by the accretion of a retrograde
globular cluster in a pre-existing disk of stars \citep{Sambhus2002}.
With the intention of exploring the outcomes of such accretion events, 
we performed a series of N-body simulations of CR stellar
distributions evolving in the sphere of influence of an SMBH. Below,
we report on results which show unstable CR distributions evolving
into stable, uniformly precessing lopsided nuclei. We follow the instability in our 
featured simulation from multiple complementary perspectives. We then probe the ensuing
lopsided nucleus, and show that in addition to displaying all the key observational signatures which Keplerian disks are meant to model, it recovers asymmetries in the tail of Adromeda's line of sight velocity distribution \citep{Bender2005} which a thick lopsided disk alone is unable to reproduce\citep{Peiris2003}. 


\begin{figure} 
\includegraphics[scale=0.65]{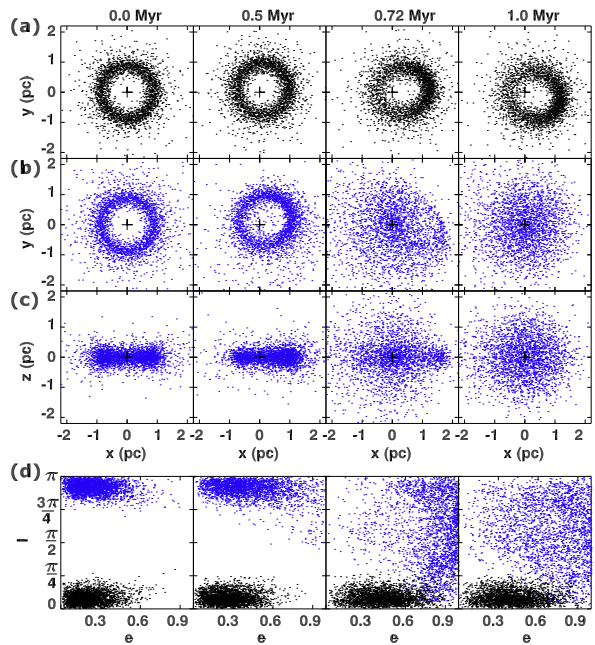} \caption{The
eccentricity-inclination dynamics of
  the particles up to 1 Myr is shown. In row {\bf a}, we follow a top
  view
  of the prograde particles, from initial axisymmetric annular
  configuration,
  through a phase of growth of lopsidedness, with growing mean
  eccentricity
  (0.41 around 1 Myrs), and alignment of apsides. Row {\bf b} shows a
  top
  view of retrograde particles over the same period; they experience
  greater
  increase in mean eccentricity (around 0.73 by 0.72 Myrs), and show
  signs
  of apse alignment by 0.5 Myrs, before dispersing (by 1 Myrs) into a
  disky
  blob. A side view of retrograde particles in row {\bf c} shows how
  the
  in-plane dispersal seen in row {\bf b} reflects in projection a
  burst of
  out-plane dynamics which puffs the initially retrograde disk into a
  triaxial cluster of stars. The eccentricity-inclination dynamics of the
  full
  cluster is captured in row {\bf d}, with prograde in black, and
  retrograde
  in blue; the mean eccentricity of both populations grows as the
  instability
  develops; by 1 Myrs, the mean inclination of the prograde particles
  has
  hardly changed, while the highly eccentric retrograde population, is
  widely
  dispersed in inclination.}  \label{fig:part} \end{figure}

\section[]{Counter-Rotating Instability: From Linear Growth to
Saturation}

Our N-body simulations were performed with the parallel version of
\gadget~\citep{Springel2005}, an oct-tree code, which is popular with
the cosmological structure formation community. We pushed this tool to
the limit of extreme mass ratios, one for which it was not
specifically designed, one in which it did the desired job, albeit
with stringent force accuracy, and time stepping criteria. With M31 in
mind, all experiments have an SMBH with a mass of \msmbh=10$^{8}~$\SM
and a main disk component (prograde in our convention) with a tenth of
SMBH mass \citep{Bender2005, Peiris2003}. An exhaustive exploration,
with sufficient realism, of a scenario in which a CR cluster
($10^5-10^6$ \SM) decays under dynamical friction to nuclear regions
\citep{tremaine1975}, then gets disrupted \citep{Kim2003, Fujii2010}
as it couples to the massive nuclear stellar disk, would have been
forbidding to pursue in detail with our current computational
resources. Instead we opted for an extensive exploration of a
realistic configuration which is known to be unstable \citep{Touma2002,
Sridhar2010}, and in which the disrupted CR cluster is modeled by
overlaying the prograde disk with a retrograde disk of 1/10 its
mass. Details on initial conditions for this experiment, along with
\gadget~simulation parameters, are described in Appendix
\ref{sec:supp-methods}; variations on the fiducial experiment are
discussed in Appendix \ref{sec:crExplored}.

\subsection{Dissecting the Instability} CR configurations are known to
be (linearly) unstable \citep{Touma2002, Sridhar2010}.  We are here
concerned with the (non-linear) saturation of this instability and
shall first map its unfolding with complementary views of particle
dynamics (Fig. \ref{fig:part}). A top view of stars, in both prograde
and retrograde populations (hereafter PP and RP respectively), shows
them developing lopsidedness by coupling growing eccentricity to
apse-alignment. The more massive PP maintains its coherent,
apse-aligned precession (Fig. \ref{fig:part},{\bf (a)}) whereas the
lighter RP dissolves gradually into a disky blob after 0.5 Myrs
(Fig. \ref{fig:part},{\bf (b)}). Viewed from the side, the gradual
in-plane dispersal of the RP occurs along with a dramatic excitation
of out-of-plane motion; by 1 Myrs the RP unfolds into a triaxial structure
(Fig. \ref{fig:part},{\bf (c)}). Scatter plots of particle
eccentricities and inclinations (Fig. \ref{fig:part},{\bf (d)}) reveal
how, by 0.72 Myrs, the now highly eccentric RP is well on its way to
complete triaxial disruption. Tucked in this emerging cluster, the PP,
which has absorbed much of the RP's (negative) angular momentum, heats
up slightly in inclination as it consolidates its eccentric lump, then
adjusts slowly to whatever little angular momentum is left in the RP.

\begin{figure} \centering \includegraphics[scale=1.0]{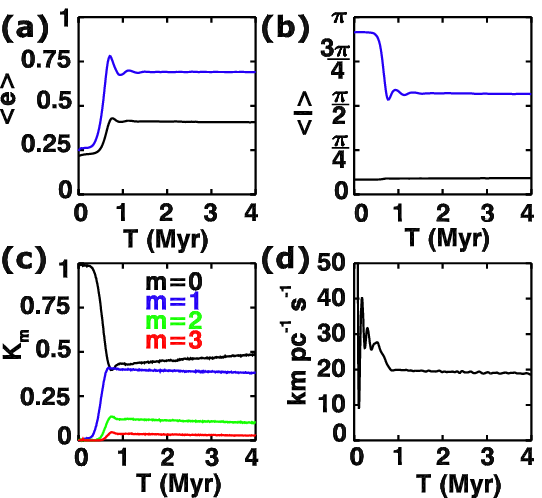}
\caption{The linearly unstable, counter-rotating
  configuration saturates into a lopsided uniformly precessing disk.
  Panels {\bf a} and {\bf b} show the mean eccentricity $<$e$>$ and
  inclination $<$I$>$ of PP (black) and RP (blue). $<$e$>$ grows, at
  near constant $<$I$>$, till about 0.5 Myrs, at which point $<$I$>$
  takes off, with both $<$e$>$ and $<$I$>$ peaking around 0.72 Myrs.
  Cycles of increasingly smaller amplitude follow this initial growth
  phase,
  leading to a saturated eccentric prograde disk which is shrouded by
  a
  triaxial halo. In panel {\bf c}, the power of the four dominant
  modes
  in projected density is displayed; an initially axisymmetric
  configuration
  (m=0) lends way to a dominant m=1 mode, along with non-negligible
  m=2
  and m=3 contributions. Triaxial dispersal of the RP
  is responsible for the slight growth of the m=0 amplitude, and the
  related slight decrease in the m=1 mode amplitude after 0.72
  Myrs. In panel {\bf d},
  we follow the precession rate of the m=1 mode in time: initial large
  amplitude oscillations reflect back and forth libration of prograde
  and
  retrograde m=1 excitations about the uniform precession state; these
  oscillations die out with the dispersal of the retrograde bunch,
  leaving
  an m=1 mode which precesses at a near constant rate of 19 km
  s$^{-1}$ pc$^{-1}$.}  \label{fig:statvstime} \end{figure}

A look at the mean eccentricity and inclination of both populations
(Fig. \ref{fig:statvstime} {\bf a, b}) confirms these observations, as
it reveals three distinct phases: {\bf Phase-I} of near coplanar
growth of the mean eccentricity of both populations which lasts till
about 0.5 Myrs; {\bf Phase-II} of continued mean eccentricity growth,
but now coupled to growth in the mean inclination of the RP, both of
which reach a maximum around 0.72 Myr; {\bf Phase-III}, with cycles of
increasingly small amplitude leading to a near-steady regime. As
emphasized in the caption of Fig. \ref{fig:statvstime}, transitions in
mean eccentricity and inclination are neatly imprinted on the modal
structure of the projected density and the pattern speed of the
dominant m=1 mode, as the full cluster transitions from a thin
axisymmetric initial state, to a saturated, lopsided, and slowly
precessing mode (Fig. \ref{fig:statvstime}, {\bf c, d}, and
Fig. \ref{fig:dens}).

Saturation of the m=1 mode is evidently correlated with the dispersal
of the initially retrograde population into a triaxial cluster of high
eccentricity orbits. The dispersal appears to be associated with
collective dynamics along eccentricity-inclination cycles. These
cycles are excited when a population of stars (the retrograde
population) develops sufficient eccentricity (through a CR
instability) in the presence of a dominant eccentric perturber (the
more massive eccentric and precessing prograde disk)
\cite{Touma2009}. One can presumably describe the inception of these
cycles in a generalized Kozai-Lidov framework \citep{naoz2011}, but
then one is left with the harder task of accounting for the dispersal
of the population on these cycles. We gained valuable insight into
dispersal then saturation by modeling a closely related process in a
2D analog of the 3D cluster (Kazandjian \& Touma, in
preparation). The model in question permits a blow by blow account of
the mutual sculpting of planar prograde and retrograde populations in
terms of capture in, then escape from a drifting, trapping region in
phase-space \citep{sridhar1997}.  It can naturally explain the
near-coplanar growth and apse alignment in phase {\bf I} of 3D
dynamics. Furthermore, it clarifies the collisionless damping process
at work in the decaying cycles of phase {\bf III}. How the
eccentricity growth of phase {\bf I} paves the way for inclination
growth in phase {\bf II} cannot be captured in a planar analogue and
shall await a more sophisticated treatment of the 3D deployment of
unstable CR clusters.

\section{Confrontation with M31: Preliminary Results}

\begin{figure} \centering \includegraphics[scale=1.0]{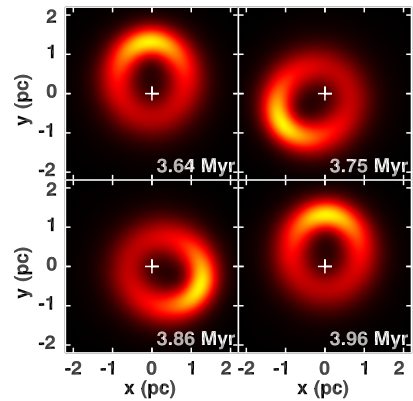}
\caption{The projected density of the saturated configuration at the
end of the simulation is followed over a precession cycle; a P1-like
peak is evident around apse-alignment; the distribution is practically
invariant as it revolves uniformly over its 0.32 Myr cycle.}
\label{fig:dens} \end{figure}

\begin{figure} \centering \includegraphics[scale=0.70]{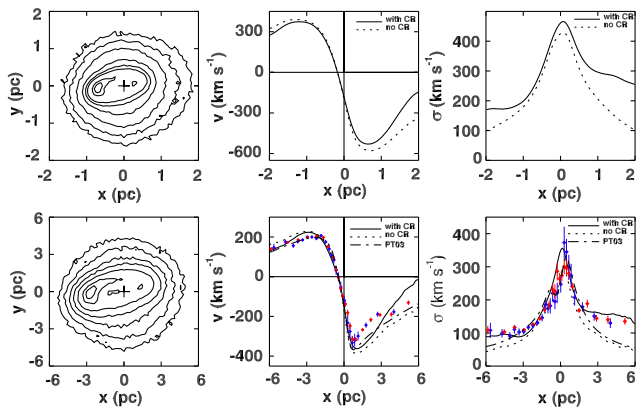}
\caption{Photometry and kinematics of saturated configurations are
probed, and a remarkable agreement with the kinematics of Andromeda's nucleus is
demonstrated. In the {\bf top row}, we show the results of photometric and
kinematic observations of our fiducial simulation: a double-peak is evident
in the surface brightness distribution{top \bf[left]}, with a brighter 
(P1-like) peak on the left of the SMBH at the origin, and a fainter (P2-like) 
peak on its right; the zero point in the rotation curve {top \bf [center]} is
shifted towards brighter peak and slower rotation; the peak velocity dispersion
is displaced towards fainter peak and faster rotation {\bf [right]}; rotation and 
LOSVD curves are shown without ({\bf dashed}) and with ({\bf solid}) contributions
from the dispersed CR disk, which contributions introduce a significant asymmetry
in the tails. The best fit (non-aligned) models of \protect\cite{Peiris2003} 
(labeled PT03) with no bulge are also displayed as {\bf dot-dashed} lines. 
A Quantitative agreement is demonstrated in the {\bf bottom row, center and right} panels, 
between a dynamically similar copy of the featured simulation
 (observed under HST conditions) ({\bf solid line}) and M31's kinematics
 \citep{Bender2005} (blue and red dots). \label{fig:kin}}
  \end{figure}

The saturated distribution of stars, which is followed in projection,
and over a full precessional cycle in Fig. \ref{fig:dens}, generalizes
Keplerian disk models of M31's nucleus \citep{Peiris2003} in two
obvious ways. On one hand, the eccentric disk in our nucleus is
shrouded by a lighter triaxial cluster, whereas currently preferred
models tend to work with the eccentric disk alone: as discussed below,
the additional triaxial cluster contributes a crucial improvement to
model kinematics. On the other hand, successful eccentric disk models
\citep{Peiris2003} are essentially kinematic in nature (their origin,
and dynamical evolution remaining uncertain), whereas ours is a
dynamically stable nucleus to which an unstable configuration
saturates. 
Given that the eccentric disk component in our saturated lopsided
nucleus is ten times more massive than the triaxial cluster, the
photometry and kinematics of this nucleus will naturally display all
the significant qualitative features which eccentric disk models seek
to explain (Fig. \ref{fig:kin}, top row): a double peak in surface
brightness, an asymmetry in the rotation velocity (with an off-center
zero velocity point), and an off-centered peak in the line of sight
velocity dispersion \citep{Bender2005, Peiris2003} (hereafter
LOSVD). Pushing our luck, we seek a preliminary quantitative
confrontation of simulation results with observations of M31's
nucleus. In so doing, we work with a rescaled and dynamically similar
version of our fiducial experiment (roughly speaking, we rescale from fiducial simulation
length scale $\simeq 1$ pc, to M31 nucleus scale, $\simeq 3$ pc).  The resulting Model Nucleus is then observed under HST conditions \citep{Bender2005, Peiris2003} (details in
Appendix \ref{sec:supp-methods}). The outcome of this procedure is then overlaid
with M31's kinematics (Fig. \ref{fig:kin}, bottom row). We closely
match the observed shift in the zero-point of rotation curve (x=-0.485
compared to x=-0.448 pc in observations \citep{Bender2005} ), the
difference in the peak rotation velocities (v=585 km s$^{-1}$ compared
to v=532 $\pm$ 41 km s$^{-1}$), the magnitude of and shift in the peak
in the LOSVD curve (x=+0.187, $\sigma$=+355.7 km s$^{-1}$ compared to
x=0.298 pc, $\sigma$=+373 $\pm$ 48 km s$^{-1}$). 

The triaxial cluster of our Model endows its kinematics
with the marked asymmetry between the tails of the LOSVD, which is clearly
observed in HST data \citep{Kormendy1999, Bender2005}, and which has
proven difficult to recover with eccentric disks alone
\citep{Peiris2003} (compare PT03 and the ``no CR'' curves to the solid
curve in the rightmost column of Fig. \ref{fig:kin}). Recall that this
cluster consists of eccentric orbits which result from the triaxial
dispersal of the initially retrograde cluster (rows {\bf b} and {\bf d} of
Fig. \ref{fig:part}). When superposed over the more massive lopsided
disk, stars in this cluster will leave their kinematic signature
mostly in the outer parts of the nucleus, which is where they linger
the most on their eccentric orbits. That signature is particularly
pronounced where the lopsided prograde disk contributes least, i.e. on
the anti-P1 side of the nucleus, hence the resulting asymmetry. 

The initial ring-like configuration in our experiment is clearly deficient
in stars in its central region. Hence, we did not expect for the resulting 
double nucleus to closely approximate the observed surface brightness
profile in that region. This said, we were pleased to learn that a
surface brightness profile, along a slit extending from brighter to
fainter peak, has maxima around 13.3 and 14.2 mag.arcsec$^{-2}$
respectively, which compare favorably with the 13.2 and 13.57
mag.arcsec$^{-2}$, observed by HST.

Last but not least, we note that for the adopted similarity transformation\footnote{In this transformation, time is rescaled by $s^{\frac{3}{2}}$,
whenever length is rescaled by $s$. The adopted $s\simeq 2.7$ accounts for the slower
precession rate of the Model Nucleus.}  the pattern speed of the lopsided mode scales down from
the 19 km s$^{-1}$ pc$^{-1}$ of the fiducial nucleus
(Fig. \ref{fig:statvstime}, {\bf d}), to the slower 4.3 km s$^{-1}$
pc$^{-1}$ of the Model Nucleus. At this relatively slow pattern speed,
gravitational perturbations by the Model Nucleus are expected to drive
and confine gas in a disk close to the SMBH \citep{Chang2007}.  When
fueled by stellar mass loss from the nucleus, such a disk can
regenerate starbursts at a rate which is consistent with the observed
compact cluster of early type stars (P3) around M31's SMBH
\citep{Bender2005, Chang2007}. Thus, M31's P3 could very well be a
direct consequence of P1 and P2 in our dynamical Model.

\section{Discussion}
Andromeda's kinematics shows a marked asymmetry between the tails of its LOSVD which is clearly unaccounted for in Keplerian disk models, and appears to be cleanly resolved by combining an eccentric disk with a triaxial cluster of highly eccentric orbits.  To be sure, thorough modeling is called for before we can rigorously quantify the improvements of a disk-cluster combination over the thick eccentric disk model of \cite{Peiris2003}. Still, we find it remarkable that this combination achieves, with minimal probing, the level of qualitative and quantitative agreement described above, and to see it emerge as a natural end product of our process can only strengthen the case for CR  stimuli of  double nuclei \citep{Lauer2005}. That case is made stronger when one  learns that a rich variety of CR configurations are equally prone to developing lopsidedness (details in Appendix \ref{sec:crExplored}.). Experiments with CR point masses suggest that a cluster, initially on a circular trajectory, can drive a coplanar disk unstable provided its orbital radius is less than a critical radius (which is roughly
equal to twice the mean radius of the disk); clusters on highly
inclined circular trajectories can still drive a CR disk into a
lopsided state; on the other hand, clusters with too large an initial
eccentricity will end up with little negative angular momentum to
drive a coplanar disk lopsided. The CR perturbation envisaged in this
work is delivered by a CR cluster which spirals deep into the nucleus
by dynamical friction with bulge stars \citep{tremaine1975} before it
disrupts in the combined tidal field of bulge and SMBH
\citep{Quillen2003}. The orbital radius at which the cluster disrupts
depends critically on the cluster's core density. For a cluster
migrating on a near-circular trajectory into M31's nucleus, we
estimate \citep{Quillen2003} that a core density $\geq$ 5 $\times$
10$^5$ \SM pc$^{-3}$ is sufficient for the cluster to cross past the
critical radius for instability. Mass segregation and/or
intermediate-mass black hole formation \citep{Kim2003, Fujii2010} can
amplify core density significantly, thus further delaying disruption
till the cluster migrates into the overlapping configurations explored
here. Such configurations may also arise when a cluster penetrates the
nucleus on an eccentric orbit, and disrupts upon close encounter with
the SMBH. CR clusters will likely approach a pre-existing nuclear disk
on eccentric, and inclined trajectories; they may thus find themselves
on the eccentricity-inclination cycles observed above, then disperse
as they drive the disk in a saturated lopsided configuration. 
Given on one hand the strong likelihood of counter-rotating excitations in 
galactic centers \citep{Jog2009}, and on the other the robustness and
efficiency of the proposed mechanism,  double nuclei are likely to 
prove ubiquitous in stellar clusters dominated by 
super-massive black holes \citep{Lauer2005}.  More generally (and irrespective
of whether or not a given lopsided nucleus results from a CR instability), the proposed mechanism can be deployed to customize triaxial equilibrium
configurations with which to model observed kinematics of stellar
black hole nuclei \citep{alexander2005PhR} (the Milky Way's included) and
improve estimates of the mass and feeding rate \footnote{We note in passing that, as it drives a substantial fraction of stars to near radial orbits, the CR instability populates the loss cone of the SMBH \citep{Magorrian99}. Repeated CR accretion events will thus enhance the feeding rate of the central SMBH as they sculpt stellar distributions in its sphere of influence.} of the black hole within \citep{Magorrian98, kormendy2004}. Whether exploring realistic CR scenarios or tailoring triaxial equilibrium models, extensive numerical simulations of CR stellar systems with state of the art solvers \citep{Fujii2010, Touma2009} will surely contribute invaluable insights into the dynamics and structure of stellar black hole nuclei. 

\section*{Acknowledgments}

 M.V.K. expresses his gratitude to the faculty, staff, and students of
 the department of physics at the American University of Beirut (AUB),
 where the bulk of this work was conducted. Computations were
 performed the Ibnsina cluster at the Center of Advanced Mathematical
 Sciences(AUB), and on the Aurora cluster at the Institute of Advanced
 Studies (Princeton). Fruitful diskussions with S. Sridhar, Scott
 Tremaine and Jarle Brinchmann are gratefully acknowledged. 
The support of NSF grants AST-0206038 and AST-0507401 was 
invaluable during all phases of this work.

\bibliographystyle{mn2e} \bibliography{references}

\begin{thebibliography}{37}
\expandafter\ifx\csname natexlab\endcsname\relax\def\natexlab#1{#1}\fi

\bibitem[{{Alexander}(2005)}]{alexander2005PhR}
{Alexander} T., 2005, \physrep, 419, 65

\bibitem[{{Bacon} {et~al}\mbox{.}(2001){Bacon}, {Emsellem}, {Combes}, {Copin},
  {Monnet}, \& {Martin}}]{Bacon2001}
{Bacon} R., {Emsellem} E., {Combes} F., {Copin} Y., {Monnet} G., {Martin} P.,
  2001, \aap, 371, 409

\bibitem[{{Bender} {et~al}\mbox{.}(2005){Bender}, {Kormendy}, {Bower}, {Green},
  {Thomas}, {Danks}, {Gull}, {Hutchings}, {Joseph}, {Kaiser}, {Lauer},
  {Nelson}, {Richstone}, {Weistrop}, \& {Woodgate}}]{Bender2005}
{Bender} R. {et~al.}, 2005, \apj, 631, 280

\bibitem[{{Chang} {et~al}\mbox{.}(2007){Chang}, {Murray-Clay}, {Chiang}, \&
  {Quataert}}]{Chang2007}
{Chang} P., {Murray-Clay} R., {Chiang} E., {Quataert} E., 2007, \apj, 668, 236

\bibitem[{{Dressler} \& {Richstone}(1988)}]{Dressler1988}
{Dressler} A., {Richstone} D.~O., 1988, \apj, 324, 701

\bibitem[{{Emsellem} \& {Combes}(1997)}]{Emsellem1997}
{Emsellem} E., {Combes} F., 1997, \aap, 323, 674

\bibitem[{{Fujii} {et~al}\mbox{.}(2010){Fujii}, {Iwasawa}, {Funato}, \&
  {Makino}}]{Fujii2010}
{Fujii} M., {Iwasawa} M., {Funato} Y., {Makino} J., 2010, ArXiv e-prints

\bibitem[{{Gerhard}(2001)}]{Gerhard2001}
{Gerhard} O., 2001, \apjl, 546, L39

\bibitem[{{Gultekin} {et~al}\mbox{.}(2011){Gultekin}, {Richstone}, {Gebhardt},
  {Faber}, {Lauer}, {Bender}, {Kormendy}, \& {Pinkney}}]{Gultekin2011}
{Gultekin} K., {Richstone} D.~O., {Gebhardt} K., {Faber} S.~M., {Lauer} T.~R.,
  {Bender} R., {Kormendy} J., {Pinkney} J., 2011, ArXiv e-prints

\bibitem[{{Hopkins} \& {Quataert}(2010{\natexlab{a}})}]{Hopkins2010a}
{Hopkins} P.~F., {Quataert} E., 2010{\natexlab{a}}, \mnras, 407, 1529

\bibitem[{{Hopkins} \& {Quataert}(2010{\natexlab{b}})}]{Hopkins2010b}
{Hopkins} P.~F., {Quataert} E., 2010{\natexlab{b}}, \mnras, 405, L41

\bibitem[{{Jacobs} \& {Sellwood}(2001)}]{Jacobs2001}
{Jacobs} V., {Sellwood} J.~A., 2001, \apjl, 555, L25

\bibitem[{{Jog} \& {Combes}(2009)}]{Jog2009}
{Jog} C.~J., {Combes} F., 2009, \physrep, 471, 75

\bibitem[{{Kim} \& {Morris}(2003)}]{Kim2003}
{Kim} S.~S., {Morris} M., 2003, \apj, 597, 312

\bibitem[{{Kormendy}(1988)}]{Kormendy1988}
{Kormendy} J., 1988, \apj, 325, 128

\bibitem[{{Kormendy}(2004)}]{kormendy2004}
{Kormendy} J., 2004, Coevolution of Black Holes and Galaxies, 1

\bibitem[{{Kormendy} \& {Bender}(1999)}]{Kormendy1999}
{Kormendy} J., {Bender} R., 1999, \apj, 522, 772

\bibitem[{{Lauer} {et~al}\mbox{.}(2005){Lauer}, {Faber}, {Gebhardt},
  {Richstone}, {Tremaine}, {Ajhar}, {Aller}, {Bender}, {Dressler},
  {Filippenko}, {Green}, {Grillmair}, {Ho}, {Kormendy}, {Magorrian}, {Pinkney},
  \& {Siopis}}]{Lauer2005}
{Lauer} T.~R. {et~al.}, 2005, \aj, 129, 2138

\bibitem[{{Lauer} {et~al}\mbox{.}(1993){Lauer}, {Faber}, {Groth}, {Shaya},
  {Campbell}, {Code}, {Currie}, {Baum}, {Ewald}, {Hester}, {Holtzman},
  {Kristian}, {Light}, {Ligynds}, {O'Neil}, \& {Westphal}}]{Lauer1993}
{Lauer} T.~R. {et~al.}, 1993, \aj, 106, 1436

\bibitem[{{Lauer} {et~al}\mbox{.}(1996){Lauer}, {Tremaine}, {Ajhar}, {Bender},
  {Dressler}, {Faber}, {Gebhardt}, {Grillmair}, {Kormendy}, \&
  {Richstone}}]{Lauer1996}
{Lauer} T.~R. {et~al.}, 1996, \apjl, 471, L79+

\bibitem[{{Light} {et~al}\mbox{.}(1974){Light}, {Danielson}, \&
  {Schwarzschild}}]{Light1974}
{Light} E.~S., {Danielson} R.~E., {Schwarzschild} M., 1974, \apj, 194, 257

\bibitem[{{Magorrian} \& {Tremaine}(1999)}]{Magorrian99}
{Magorrian} J., {Tremaine} S., 1999, \mnras, 309, 447

\bibitem[{{Magorrian} {et~al}\mbox{.}(1998){Magorrian}, {Tremaine},
  {Richstone}, {Bender}, {Bower}, {Dressler}, {Faber}, {Gebhardt}, {Green},
  {Grillmair}, {Kormendy}, \& {Lauer}}]{Magorrian98}
{Magorrian} J. {et~al.}, 1998, \aj, 115, 2285

\bibitem[{{Naoz} {et~al}\mbox{.}(2011){Naoz}, {Farr}, {Lithwick}, {Rasio}, \&
  {Teyssandier}}]{naoz2011}
{Naoz} S., {Farr} W.~M., {Lithwick} Y., {Rasio} F.~A., {Teyssandier} J., 2011,
  ArXiv e-prints

\bibitem[{{Peiris} \& {Tremaine}(2003)}]{Peiris2003}
{Peiris} H.~V., {Tremaine} S., 2003, \apj, 599, 237

\bibitem[{{Quillen} \& {Hubbard}(2003)}]{Quillen2003}
{Quillen} A.~C., {Hubbard} A., 2003, \apj, 125, 2998

\bibitem[{{Salow} \& {Statler}(2001)}]{Salow2001}
{Salow} R.~M., {Statler} T.~S., 2001, \apjl, 551, L49

\bibitem[{{Sambhus} \& {Sridhar}(2002)}]{Sambhus2002}
{Sambhus} N., {Sridhar} S., 2002, \aap, 388, 766

\bibitem[{{Silk}(2011)}]{Silk2011}
{Silk} J., 2011, ArXiv e-prints

\bibitem[{{Springel}(2005)}]{Springel2005}
{Springel} V., 2005, \mnras, 364, 1105

\bibitem[{{Sridhar} \& {Saini}(2010)}]{Sridhar2010}
{Sridhar} S., {Saini} T.~D., 2010, \mnras, 404, 527

\bibitem[{{Sridhar} \& {Touma}(1997)}]{sridhar1997}
{Sridhar} S., {Touma} J., 1997, \mnras, 292, 657

\bibitem[{{Touma}(2002)}]{Touma2002}
{Touma} J.~R., 2002, \mnras, 333, 583

\bibitem[{{Touma} {et~al}\mbox{.}(2009){Touma}, {Tremaine}, \&
  {Kazandjian}}]{Touma2009}
{Touma} J.~R., {Tremaine} S., {Kazandjian} M.~V., 2009, \mnras, 394, 1085

\bibitem[{{Tremaine}(1995)}]{Tremaine1995}
{Tremaine} S., 1995, \aj, 110, 628

\bibitem[{{Tremaine}(2001)}]{Tremaine2001}
{Tremaine} S., 2001, \aj, 121, 1776

\bibitem[{{Tremaine} {et~al}\mbox{.}(1975){Tremaine}, {Ostriker}, \&
  {Spitzer}}]{tremaine1975}
{Tremaine} S.~D., {Ostriker} J.~P., {Spitzer}, Jr. L., 1975, \apj, 196, 407

\end{thebibliography}


\setcounter{section}{0}
\renewcommand \thesection{\Alph{section}}
\renewcommand \thesubsection{\thesection.\arabic{subsection}}
\renewcommand \thesubsubsection{\thesubsection.\arabic{subsection}}

\renewcommand{\theequation}{\thesection-\arabic{equation}}
\setcounter{equation}{0}

\section[]{Methods} \label{sec:supp-methods}

We performed our simulations using the cosmological numerical tool
\gadget~\citep{Springel2005}. Forces were evaluated using the oct-tree;  
cosmological, SPH, and grid capabilities were switched off. Although this
narrowed down the parameter space significantly, one had to be careful
fine tuning parameters which control tree walk and time stepping. The
depth of the tree to be explored is controlled by \gadget's
parameter \verb|ErrTolForceAcc|: daughter cells are opened
until a certain criterion, set by \verb|ErrTolForceAcc|, is satisfied,
thus terminating the tree walk (we refer the reader to \gadget's
user guide for more details about this criterion). Once the
accelerations of all the particles are at our disposal, the positions
of these particles are advanced with a time-step $\Delta t = \sqrt{2
\eta b / {\bf |a|}} $, where ${\bf |a|}$, $\eta$ and $\epsilon$ are
the magnitude of the computed acceleration, \verb|ErrTolIntAccuracy| 
(which controls the step-size) and the gravitational softening respectively.
Setting \verb|ErrTolForceAcc| to 0.001 and \verb|ErrTolIntAccuracy| to 0.001
assured accurate enough simulations for our purposes. 

Our featured simulations consist of a massive prograde disk, on which
a light retrograde disk is overlayed, with both residing in the sphere
of influence of a central SMBH. Such a configuration is a likely
outcome of the disruption of clusters that can migrate deep in the
galactic centers \citep{Gerhard2001, Kim2003, Fujii2010}. Initial
positions and velocities of disk particles are picked from
radial and vertical distributions which are standard to disk dynamics
simulations \citep{Emsellem1997}. Since \gadget $~$has difficulties in dealing
with  high mass density contrast, we further chose
our disk to be annular in-order to avoid close encounters with the
SMBH. Thus, we introduced an inner and outer radial cut-off when
sampling the disks. The SMBH is modeled as a Plummer-softened particle
with softening length of b=0.01 pc. To minimize sources of instability
which are not related to counter-rotation, disk components were separately
virialized before coupling them to each other or to perturbers. In the
course of virializing, the more massive disks (prograde in our
simulations) underwent short-lived gravitational instabilities, before
relaxing to a hotter near-axisymmetric equilibrium state.

The central SMBH with a mass of 10$^8~$\SM is ten times more massive
that the pre-existing disk of stars, itself ten times more massive than
the disturbing CR disk. the mean eccentricities of the virialized prograde
 and retrograde disk were $<$$e_p$$>$ = 0.22 and $<$$e_r$$>$ = 0.25 respectively,
 with corresponding standard deviations $\sigma_\mathrm{p} = $ 0.12 and $\sigma_\mathrm{r} = $
0.14. In both disks, the mean of the semi-major axis distribution was
$<$a$>$$\sim$1.06 pc with a standard deviation of $\sim$0.3 pc whereas the
 spread in inclination was around 8$^\circ$.

Simulation particles revolve around the central SMBH on perturbed
Keplerian ellipses; associated osculating orbital elements [namely
semi-major axis (a), eccentricity (e), inclination (I), argument of
periapse (g), and longitude of the node (h)] are simply recovered from
particle position and velocity. In addition to particle orbit
elements, our dynamical analysis required the time evolving Fourier
modes of the surface (projected) density of the CR
cluster, and associated potentials.  The surface density $\Sigma(r,
\theta, t)$ is Fourier expanded in $\theta$:

\begin{equation} 
\label{eqn:fftSum3}
\Sigma (r,\theta, t) =  a_0(r,t) + \sum \limits_{m=1}^{m=\infty} a_m(r, t) 
\cos(m\theta + \phi_m(r, t) ) 
\end{equation}
with $[a_m(r, t), \phi_m(r,t)]$, the m$^\mathrm{th}$ mode amplitude and phase respectively.
This is practically performed with FFTs over a grid in $\theta$ at various 
judiciously chosen radii. The relative power $K_m$ in a given mode m is then computed
\begin{equation}
\label{eqn:modeAmp}
  K_m (t) \equiv \frac{\int \limits_0^\infty a^2_m(r,t) 2 \pi r dr}{\sum \limits_{m=0}^{m=\infty} 
    \int \limits_0^\infty a^2_m(r,t) 2 \pi r dr},
\end{equation}

Cluster kinematics were determined from a snapshot of our benchmark 3D
simulation at T=4 Myr (well into the saturated regime). The scale length
of that simulation is on the order of 1 pc, whereas that of M31's
nucleus is more like 3 pc. To compare with M31 observations, we rescaled 
positions (and with them the softening length) by a factor $s$, and time
 by a factor $s^{\frac{3}{2}}$; velocities are naturally rescaled by
$s^{-\frac{1}{2}}$ to leave equations of motion in the new variables
invariant. The adopted scaling ($s \simeq 2.7$) was first constrained
within the interval $ 2.0 < s < 3.1 $, by matching peaks in the
rotation curve, then fine tuned to fit the kinematics, while adjusting
the sky angles as well.

While searching for the optimal sky angles \citep{Peiris2003} of our rescaled snapshot, we
took into account artifacts of the HST SITS instrument by convolving
all the particles with a double Gaussian \citep{Bender2005}. The slit width was
set to $0.1''$ with a PA of $39^{\circ}$. Both the fiducial, and
rescaled snapshots were observed with the same sky angles $\theta_a =
-59.1^{\circ}$, $\theta_i = 65.3^{\circ}$. We note that larger $\theta_i$ 
permits an almost perfect match with the observed LOSVD (remarkably so 
for $\theta_i$ close to M31's inclination of 77$^{\circ}$), with deteriorating
 rotation curve; while smaller $\theta_i$ improved the fit to the rotation 
curve dramatically while disturbing agreement in the LOSVD.


\section[]{The CR Instability Explored} \label{sec:crExplored}

  Here, we summarize results of a battery of experiments conducted
  with CR configurations that are likely to occur in
  galactic centers.  
\\First, we explore dynamics in highly inclined
  configurations, by considering an extreme situation: a lower
  resolution ( 50$\times$10$^3$ particles ) analogue of the 3D
  benchmark simulation was evolved with equal mass annuli (
  M$_r$=M$_p$=0.05 \msmbh, M$_r$ and M$_p$ are the masses of the retrograde
  and prograde disks respectively) where the PP and RP disks
  have a relative inclination of 90$^\circ$.  This experiment was 
  motivated by our concern about non-coplanar, CR rings 
  in galactic centers, and   whether such configuration may result in m=1
  instabilities. The configuration which was violently unstable, resulted
  in both components merging, then relaxing into a thick, precessing, and
  lopsided disk.
\\Second, we  explored the effect of an in-spiraling cluster on our
  annular disk \citep{Fujii2010}. Evolving an actual realization of a plummer sphere 
  with \gadget$~$ around the SMBH with enough accuracy is prohibitively
  expensive, thus the cluster was modeled as a point mass with a large softening 
  radius 0.1 pc. The point mass was put on a retrograde Keplerian orbit 
  around the SMBH. The massive disk of our 3D simulation was used as the prograde 
  component (M$_p$=0.1 \msmbh, $<$a$>$=0.92 pc), but 
  with 25$\times$10$^3$ particles instead of 250$\times$10$^3$. 
  The default values of the point mass (cluster) orbit are 
  M$_{\mathrm{c}}$=0.01 \msmbh$~$ (M$_{\mathrm{c}}$ is the mass of the cluster),
  semi-major axis a$_{\mathrm{c}}$=0.9 pc, eccentricity e$_{\mathrm{c}}$=0, 
  and inclination I$_{\mathrm{c}}$=$180^{\circ}$.
  The simulations for this scenario consisted of four suites. 
  In each suite one parameter was varied while keeping the remaining
  parameters to their default value:
\begin{itemize}
\item {\bf Suite 1.} $\mathrm{M}_{\mathrm{c}}$ was varied from 0.001 to 0.1 \msmbh:
  The prograde disk developed an m=1 mode for $\mathrm{M}_{\mathrm{c}} > $ 0.01 \msmbh.

\item {\bf Suite 2.} The inclination of cluster was varied from
  90$^{\circ}$ to 180$^{\circ}$: In all cases the cluster and the disk
  ended up in the same plane, and with an m=1 growing instability mode
  when the relative inclination was larger than 90$^\circ$.

\item {\bf Suite 3.} The eccentricity was varied from 0 to 0.9: the
  low eccentricity m=1 growing instability shuts off for $e_{\mathrm{c}}>0.6$.

\item {\bf Suite 4.} The semi-major axis was varied from 0.8 to 10 pc: for $\mathrm{a}_{\mathrm{c}} > 2$$<$a$>$
  the annulus remained axisymmetric, with an m=1 modes growing for
  $\mathrm{a}_{\mathrm{c}} < 2$$<$a$>$  
\end{itemize}


\label{lastpage}

\end{document}